\documentclass[epj,referee]{svjour}
\usepackage{graphics}
\usepackage{amsmath, amssymb}
\begin{document}
\title{Statistical equilibrium in simple exchange games I}
\subtitle{Methods of
solution and application to the Bennati-Dragulescu-Yakovenko (BDY) game}
\author{Enrico Scalas\inst{1} \and Ubaldo Garibaldi\inst{2} \and Stefania Donadio\inst{3}
}                     
%
%
\institute{Department of Advanced Sciences and Technology, Laboratory on Complex
Systems, East Piedmont University, Via Bellini 25 g, 15100 Alessandria,
Italy \and IMEM-CNR, Physics Department, Genoa University, via Dodecaneso 33, 16146,
Genoa, Italy \and INFN, Physics Department, Genoa University, via Dodecaneso 33, 16146, Genoa,
Italy}
\date{Received: date / Revised version: date}
%
\abstract{Simple stochastic exchange games are based on random allocation of finite 
resources. These games are Markov chains that can be studied 
either analytically or by Monte Carlo simulations.
In particular, the equilibrium distribution can be derived either by
direct diagonalization of the transition matrix, or using the detailed
balance equation, or by Monte Carlo estimates. In this paper, these
methods are introduced and applied to the Bennati-Dragulescu-Yakovenko (BDY) game.
The exact analysis shows that the statistical-mechanical analogies
used in the previous literature have to be revised.} 
\maketitle

\section{Introduction}
Agent-based models used for simulating
the allocation of finite resources in economics include
$g$ agents that can interact. These interactions can be direct and can 
include both two-body and many-body terms, but they can also be indirect, through 
some coupling and feedback mechanism with an external "field".

Each agent $i$ is characterized by a certain quantity $n_i$, which represents either size, 
or wealth or another relevant quantity. The interactions determine a variation of $n_i$ 
as a function of time. In the models, the evolution of the system can be described 
both in continuous time and in discrete time. 
In this framework, it is worth mentioning the so-called Interacting Particle Systems paradigm 
that includes, as special cases, percolation, the Ising model, the voter model, and the 
contact model \cite{liggett85}.

In general, these models are Markov chains or con\-ti\-nuous-time Markov processes. Therefore, there 
is a full set of mathematical tools to analyze them and compute the equilibrium distribution.
In this paper, however, the focus is on conservative models, where the total number of agents,
$g$, and the total size or wealth, $n=\sum_{i=1}^{g} n_i$, are conserved by the dynamics.

John Angle has introduced the so-called One Parameter Inequality Process (OPIP) that can be
defined as follows. Let us suppose that there are $g$ players in a room, each of them with an 
initial amount of money, $n_{i} (0) = n/g$. Two individuals are randomly selected to play against 
each other. They flip a coin and the winner gets a fixed fraction, $\omega$, of the 
loser's money. Then the game is iterated. If $j$ and $k$ are the selected players at step $t$, 
their amount of money at step $t+1$ is given by: 
$$
n_{j} (t+1) =  n_{j} (t) + \omega d(t+1) n_{k} (t) - \omega (1 - d(t+1)) n_{j} (t),
$$
and
$$
n_{k} (t+1) =  n_{k} (t) - \omega d(t+1) n_{k} (t) + \omega (1 - d(t+1)) n_{j} (t),
$$
where $d(t)$ is a Bernoullian random variable assuming the value $1$ with
probability $1/2$ or the value $0$ with probability $1/2$. Angle has studied the equilibrium
distribution for the OPIP by means of Monte Carlo simulations and analytical approximations
\cite{angle86,angle96,angle02}.

The Bennati-Dragulescu-Yakovenko (BDY) model described in \cite{bennati88,bennati93}
and rediscovered in \cite{dragulescu00} is very similar to the OPIP, 
but there is an important difference. After the coin toss, the winner receives a fixed amount 
of money, $d$. Indebtedness is impossible: Players reaching $n_i = 0$ cannot lose money any more. 
If they are selected to play and they lose, they stay with no money, if they win, they get the 
fixed amount of money from the loser. 
On the contrary, in the OPIP, very poor agents always lose only a fraction of their money, 
and they never reach the situation $n_i = 0$. In the OPIP, the variables $n_i$ are
intrinsically continuous, whereas in the BDY model they can be considered discrete.

To summarize, the BDY game can be described as follows. Let us consider a system of $g>1$
individuals (agents) who share $n$ coins, $n \geq g$. At each discrete time
step two agents are chosen, and they toss a coin. At
the end of the bet, the winner has one more coin and the loser has one coin
less ($d=1$ is assumed, without loss of generality). Agents' choice is random (i.e. each distinct couple has the same
probability to be extracted) and each bet is fair. If the
loser has no coins, then the move is forbidden and a new couple of players is
extracted. 
An equivalent formulation of the game, avoiding forbidden moves, is the following.
An agent is chosen randomly among all those having at least one coin, and
this agent is declared to be the loser; the winner is chosen randomly
among all agents. 

This paper will be devoted to an analysis of the BDY game. In section II, 
the basic random variables for the description of the game will be introduced.
Section III will be devoted to the methods of solution and it is the core section of this
paper. Finally, in section IV a critical discussion of the results will be presented.
The reader will find further mathematical details in an appendix.

\section{Random Variables}
In the BDY game, as well as in similar exchange games, one has to allocate
$n$ coins among $g$ agents. In the following, a random variable will be denoted
by a capital letter: $A$, whereas $a$ will refer to a specific value or
{\em realization}.

The most complete description of the game states is in terms
of {\em coin configurations}: $%
\mathbf{X}=(X_{1},\ldots ,X_{n})$. Each random variable $%
X_{i}$ is associated to the $ith-$coin, and its
range is the set of agents; for instance, $X_{7}=3$ denotes that the $7th-$coin
belongs to the $3rd-$agent. The total number of configurations for $n$ coins
distributed among $g$ agents is $g^{n}$. This can be called the
{\em coin description}.

The second (and most important in the present case) description 
is in terms of {\em coin occupation numbers}, $\mathbf{Y}=(Y_{1},\ldots ,Y_{g}),$ where the random
variable $Y_{j}$ denotes the number of coins (the wealth) of the $jth-$%
agent. If the set of configurations, $\mathbf{X}$, is known, then $Y_{j}|%
\mathbf{X}=\#$ $\{X_{i}:X_{i}=j,i=1,...,n\}$, that is the value of $Y_j$
conditioned on $\mathbf{X}$ is the number, $n_j$, of all $X_i$ equal to $j$. 
Then, one can define $\mathbf{Y=n}:=(n_{1},\ldots ,n_{g})$ as the set of
occupation numbers; they
satisfy the constraint $\sum_{1}^{g}n_{i}=n$.
This can be called the {\em agent description}. It tells us the
number of coins (wealth) of each agent. The total number of distinct 
agent descriptions for $g$ agents sharing $n$ coins is $\dbinom{n+g-1}{n}.$

The less complete description is in terms of {\em coin occupancy numbers}
or {\em partitions}: $\mathbf{Z}=(Z_{0}\ldots Z_{n})$,
where $Z_{h}|\mathbf{Y}=\#\{n_{j}=h,j=1,...,g\}$, that is the number (not
the names or labels) of agents with $h$ coins. This is the frequency distribution of
agents, commonly referred to as \emph{wealth distribution}; it is
an event, not to be confused with a probability distribution.
The constraints for $\mathbf{Z}$ are $\sum_{0}^{n}z_{i}=g,%
\sum_{0}^{n}iz_{i}=n$. For the BDY game, the number of agents without money, 
$z_0$, is very important. Its
complement is $k=g- z_{0}$, the number of agents with at least one coin.

\section{Methods of solution}

\subsection{An irreducible Markov chain}
The dynamic mechanism of the BDY game is the hopping of a coin from
one agent (the loser) to another (the winner). The
natural description is in terms of agents, $\mathbf{Y}=(n_{1},\ldots
,n_{g}).$ Let us suppose that at given time, $t$, the agents are described by 
the state $\mathbf{Y}%
(t)=(n_{1},\ldots ,n_{g}):=\mathbf{n}$. At the next step, the possible values
of $\mathbf{Y}(t+1)$ are: $\mathbf{Y}(t+1)=%
(n_{1},..,n_{i}-1,...,n_{j}+1,..,n_{g}):=\mathbf{n}_{i}^{j},$
corresponding to the a loss of the $ith-$ agent and a win of the $jth-$
one. The transition probability between these states is: 
\begin{equation}
P(\mathbf{n}_{i}^{j}|\mathbf{n)=}\frac{1-\delta _{n_{i},0}}{g-z_{0}(\mathbf{n%
})}\frac{1-\delta _{i,j}}{g-1}  \label{trans}
\end{equation}
where the first term, $(1-\delta _{n_{i},0})/(g-z_{0}(\mathbf{n}))=(%
1-\delta _{n_{i},0})/k(\mathbf{n})$, describes the random choice of the
loser among the agents with at least one coin ($n_{i}>0)$, and the second term, $(1-\delta
_{i,j})/(g-1)$, is the probability that the $jth-$ agent is the winner. 
As also an agent with zero coins can be a winner, there are no absorbing states. 
Note that in eq. (\ref{trans}) the assumption is made that coins
necessarily change agent; if one admits that
coins can come back to the loser, the second term
simplifies to $1/g$, the dynamics slightly changes, but the equilibrium distribution is not
affected.
Considering both the intuitive meaning of the game and the formal transition
probability (\ref{trans}), the sequence $\mathbf{Y}(0),%
\mathbf{Y}(1),...,\mathbf{Y}(t)$ is a discrete-space and discrete-time Markov
process, {\em i.e.} a finite Markov chain; every state can be reached from any other state, the
set of states is irreducible, and no periodicity is present. Hence, there exists an
invariant probability distribution, and this distribution coincides with the equilibrium one. 
This means that $\lim_{t->\infty }P(\mathbf{Y}(t)=%
\mathbf{n}|\mathbf{Y}(0)=\mathbf{n}^{\prime })=\pi (\mathbf{n),}$
independently from the initial state $\mathbf{Y}(0)=\mathbf{n}^{\prime }.$
Moreover, $\pi (\mathbf{n)>}0$ holds for all the $\dbinom{n+g-1}{n}$ possible occupation numbers.

\subsection{Direct enumeration and Monte Carlo sampling}
The direct enumeration method can be used to study the game when $g$ and $n$ are not too large.
To illustrate the method, let us consider the case $g=n=3$. The total number of agent descriptions is 
$10$:
$(0,0,3)$; $(0,3,0)$; $(3,0,0)$; $(0,1,2)$; $(1,0,2)$; $(1,2,0)$;
$(0,2,1)$; $(2,0,1)$; $(2,1,0)$; $(1,1,1)$.
The transition matrix between these states can be directly computed by using the rules of the game.
For instance, the state $(0,0,3)$ can only go into the two states $(0,1,2)$ and $(1,0,2)$ with equal 
probability $1/2$. The state $(0,1,2)$ can go into the four states $(1,0,2)$, $(0,2,1)$, $(1,1,1)$, and
$(0,0,3)$ and each final state can be reached with probability $1/4$. These considerations lead to the
definition of the following $10 \times 10$ transition matrix:
$$
\mathbf{P} = \left( \begin{array}{cccccccccc}
0 & 0 & 0 & 0 & 1/2 & 1/2 & 0 & 0 & 0 & 0 \\
0 & 0 & 0 & 0 & 0 & 0 & 1/2 & 1/2 & 0 & 0 \\
0 & 0 & 0 & 0 & 0 & 0 & 0 & 0 & 1/2 & 1/2 \\
0 & 0 & 0 & 0 & 1/6 & 1/6 & 1/6 & 1/6 & 1/6 & 1/6 \\
1/4 & 0 & 0 & 1/4 & 0 & 1/4 & 0 & 1/4 & 0 & 0 \\
1/4 & 0 & 0 & 1/4 & 1/4 & 0 & 0 & 0 & 1/4 & 0 \\
0 & 1/4 & 0 & 1/4 & 0 & 0 & 0 & 1/4 & 0 & 1/4 \\
0 & 1/4 & 0 & 1/4 & 1/4 & 0 & 1/4 & 0 & 0 & 0 \\
0 & 0 & 1/4 & 1/4 & 0 & 1/4 & 0 & 0 & 0 & 1/4 \\
0 & 0 & 1/4 & 1/4 & 0 & 0 & 1/4 & 0 & 1/4 & 0 \end{array} \right).
$$
The vector, $\mathbf{\pi}$, giving the equilibrium probability distribution can be computed 
diagonalizing $\mathbf{P}$, as its transpose, $\mathbf{\pi}^{t}$ , satisfies:
$\mathbf{\pi}^{t} \mathbf{P} =  \mathbf{\pi}^{t}$. In particular, in this case, one gets:
$
\pi (0,0,3) = \pi(0,3,0) = \pi(3,0,0) = 1/18,
$
$
\pi (1,1,1) = 1/6, 
$ and
$
\pi (0,1,2) = \pi(1,0,2) = \pi(1,2,0)=\pi(0,2,1)=\pi(2,0,1)=\pi(2,1,0) = 1/9.
$
A Monte Carlo simulation of the game can help in the case of larger systems. The simulation can
sample both the transition matrix and the equilibrium distribution. Both methods, direct enumeration
and Monte Carlo sampling, are limited by the size of the state space. However, for the
BDY game a general exact solution can be derived.

\subsection{Exact solution}
As the size of the state space is a rapidly growing function of $n$ and $g$, 
the invariant distribution can be investigated via
the detailed balance equation \cite{costantini}.

Let us consider two consecutive states: $(n_{1},..,n_{i},...$, $n_{j},..,n_{g})$ and 
$(n_{1},..,n_{i}-1,...,n_{j}+1,..,n_{g}),$ with the conditions $n_{i}>0$
and $i\neq j$ . The direct flux is given by 
$$\pi(\mathbf{n)}P(\mathbf{n}_{i}^{j}|%
\mathbf{n)}=\pi(\mathbf{n)}\frac{1}{g-z_{0}(\mathbf{n})}\frac{1}{g-1};$$ 
the inverse flux is 
$$\pi(\mathbf{n}_{i}^{j}\mathbf{)}P(\mathbf{n|n}%
_{i}^{j}\mathbf{)=}\pi(\mathbf{n}_{i}^{j}\mathbf{)}\frac{1}{g-z_{0}(\mathbf{n}%
_{i}^{j})}\frac{1}{g-1}.$$ 
The two fluxes are equal if 
$$\pi(\mathbf{n)}\frac{1}{%
g-z_{0}(\mathbf{n})}\frac{1}{g-1}=\pi(\mathbf{n}_{i}^{j}\mathbf{)}\frac{1}{%
g-z_{0}(\mathbf{n}_{i}^{j})}\frac{1}{g-1},$$ 
that is if $\pi(\mathbf{n)}\frac{1%
}{g-z_{0}(\mathbf{n})}=C$, where $C$ is a constant.

Hence the probability function: 
\begin{equation}
P(\mathbf{Y=n)=}\pi (\mathbf{n)=}Ck(\mathbf{n})=C(g-z_{0}(\mathbf{n}))
\label{inv}
\end{equation}
is invariant, and it coincides with the equilibrium one.

Two remarks are useful. First of all, in eq. (\ref{inv}) $\pi (\mathbf{n)}$ does not
depend on the agent labels but is a function of the partition $\mathbf{Z(n%
})$ to which the description $\mathbf{Y=n}$ belongs. This implies that all
the sequences $\mathbf{Y=n}^{\prime }$  and $\mathbf{Y=n}$ are equiprobable$%
\mathbf{,}$ if $\mathbf{n}^{\prime }$ and  $\mathbf{n}$ belong to the same $%
\mathbf{Z}$, that is if $\mathbf{n}^{\prime }$ is any permutation of $%
\mathbf{n.}$ Therefore, the random variables $(Y_{1},\ldots ,Y_{g})$ are
exchangeable \cite{costantini}, and they are also
equidistributed, once equilibrium has been reached and eq. (\ref{inv}) holds. 
All $\mathbf{n}$ belonging to the same $\mathbf{Z}$ being equiprobable,
one gets for the partition probability distribution: 
\begin{equation}
\Pi (\mathbf{z})=\frac{g!}{\prod_{0}^{n}z_{i}!}\pi (\mathbf{n})=C\frac{g!}{%
\prod_{0}^{n}z_{i}!}(g-z_{0}(\mathbf{n}))  \label{invz}
\end{equation}

Secondly, only those agent descriptions sharing the same number of agents
without coins have the same probability. Indeed, the probability of a given
occupation vector $\mathbf{n}$ depends on $z_{0}(\mathbf{n}),$ and, thus, it is
not uniform. The reader is invited to verify
this property in the particular case $g=n=3$ described in the previous subsection. 

The hypothesis of equal \emph{a priori} probabilities for all the agent
descriptions seems at the basis of Bennati's and Dragulescu and Yakovenko's
analysis of the game, whose conclusions are not fully correct if one
considers eq. (\ref{inv}). This hypothesis on occupation numbers can already be
found in a paper by Boltzmann published in 1868 and leading
to the so-called {\em Bose-Einstein} statistics 
\cite{boltzmann68,bach90,costantini97}. Indeed, if $\pi(\mathbf{n})$ were
uniform in eq. (\ref{invz}), one would get the most
probable value of $z_i$ by maximizing the multinomial
prefactor subject to the constraints for $\mathbf{Z}$. 
In the limit of large systems, the result is
$z_{i}^{*} = \dfrac{1}{a}e^{-\frac{i}{a}}$. At the end of the next subsection,
the limit $n>>g>>1$ will be considered for the BDY game, where the exponential wealth distribution is 
recovered as an approximation to the exact solution.

The normalization constant $C$ is computed in the Appendix, based on 
the method described in a paper by Hill \cite{hill77}. 
It turns out that:
\begin{equation}
C=\dfrac{1}{\sum_{k=1}^{g} k \dbinom{g}{k}\dbinom{n-1}{k-1}}
\label{cnorm}
\end{equation}

Eqs. (\ref{inv}), (\ref{invz}), together with the normalization (\ref{cnorm}), give the
equilibrium distributions for the BDY game. 

\subsection{The average wealth distribution}
The number of the agent descriptions, $\mathbf{n}$, and of the
partitions, $\mathbf{Z}$, is very large for $g$ and $n$ large. Moreover,
both $\pi (\mathbf{n})$ and $\Pi (\mathbf{z})$ are multidimensional
distributions. In order to search for a quantity that can compared with
experimental observations, one can notice that agents are
exchangeable and any probability distribution is symmetric with
respect to the exchange of their labels. Empirical data are given in terms of
the actual wealth distribution $\mathbf{z}$. At any step, $\mathbf{Z}(t)=\mathbf{z}(t)$
is just the actual wealth distribution. If equilibrium is reached,
$\Pi (\mathbf{z})$
represents the multivariate sampling distribution, and the vector $%
E(\mathbf{z)}$ denotes the set of first moments of $\Pi (\mathbf{z})$.
It is useful to define the marginal average
\begin{equation}
E(z_{i})=\sum_{\mathbf{z}}z_{i} \Pi (\mathbf{z}).  \label{Ezi}
\end{equation}%

$\mathbf{Z}$ continuously fluctuates around $E(\mathbf{z)}$. 
As a consequence of the ergodic
thorem for Markov chains, one has
that $\lim_{t\rightarrow \infty }\frac{\sum_{s=1}^{t}z_{i}(s)}{t}%
=E(z_{i}),$ and this convergence is in probability. 
Hence, if the empirical or
simulated sequence, $\mathbf{z}(0),\mathbf{z}(1),...,\mathbf{z}(t)$, 
is available, the
comparison is possible between the time average $\frac{\sum_{s=1}^{t}z_{i}(s)}{t}$ and the
ensemble average $E(z_{i})$ predicted from the knowledge of 
$\Pi (\mathbf{z})$. $E(z_{i})$,
the average wealth distribution, will coincide with the most probable
value of $\mathbf{Z}$ (say $\mathbf{z}^{\ast}$) for large systems. 
As already noticed, if $%
\pi (\mathbf{n})$ were uniform, then one could find the most probable value 
of $\mathbf{Z}$, $\mathbf{z}%
^{\ast}$, by using Lagrange multipliers, and the functional
form of $\mathbf{z}^{\ast }$ would be exponential in the Stirling approximation. 

In the BDY game, this is not the case. However, as a consequence of eq. (\ref{inv}), $%
\pi (\mathbf{n})$ is uniform for all vectors with the same $k=g-z_0$. The exact
value of $E(z_i)$ can be derived analyzing all the agent descriptions with the same $k$. 
Conditioned on $k$,
one gets: 
\begin{equation}
\left\{ 
\begin{array}{l}
E(z_{0}|k)=g-k \\ 
\left\{ 
\begin{array}{c}
E(z_{i}|k>1)=k \dfrac{\dbinom{n-i-1}{k-2}}{\dbinom{n-1}{k-1}},i=1,...,n
\\ 
E(z_{i}|k=1)= \delta _{i,n},i=1,...,n%
\end{array}%
\right. 
\end{array}%
\right.   \label{Ezik}
\end{equation}%
and the equilibrium probability of $k$ is 
\begin{equation}
P(k)=C k \dbinom{g}{k}\dbinom{n-1}{k-1}  \label{Pik}
\end{equation}%
Finally, using eqs. (\ref{cnorm}), (\ref{Ezik}) and (\ref{Pik}), one gets 
\begin{equation}
\label{wealthdistribution}
E(z_{i})=\sum_{k=1}^{g}E(z_{i}|k)P(k),\,i=0,1,...,n
\end{equation}
The proof of the above results can be found in the Appendix. Notice that the thermodynamic
limit ($n,g,k>>1)$ of eq. (\ref{Ezik}) is $\dfrac{E(z_{i+1}|k)}{k}\simeq 
\dfrac{k}{n}\left( 1-\dfrac{k}{n}\right) ^{i}$. Then, the average fraction of
agents with at least one coin follows a geometric distribution that becomes exponential
in the continuous limit. In this limit, the average wealth distribution, $E(z_{i})$, 
(or the most probable wealth distribution $z_{i}^{\ast }$) is a
mixture of exponential distributions with mixing measure given by eq. (\ref{Pik}). 

Considering eq. (\ref{Pik}), one observes that 
$$\frac{P(k+1)}{P(k)} =
\dfrac{%
(g-k)(n-k)}{k^{2}}$$
with $\frac{P(k+1)}{P(k)}>1$ for $k<k^{\ast }=\frac{ng}{n+g}$, and $\frac{P(k+1)}{P(k)}<1$ for $%
k>k^{\ast }.$ Therefore, in the case of minimum density, $n=g$, one has that $P(k)$
is bell-shaped with flat maximum at $k^{\ast }=\frac{ng}{n+g}$ and $k^{\ast
}+1,$ as $\frac{P(k+1)}{P(k)}=1$, $k^{\ast }=\frac{g}{2}$. In the large
density limit $n>>g$, the curve is left-skewed, the maximum is very close to $%
g$, as $k^{\ast }=\frac{g}{1+g/n}\simeq g(1-g/n)$. Furthermore, if $g(1-g/n)>g-1$,
i.e. $g^{2}<n$, the maximum value is just $k^{\ast }=g$. In the
case of large density, the mixing probability distribution is concentrated on a small number of
values of $k,$ and,
thus, if $g>>1$ the behaviour is not very different from the single geometric distribution
$\dfrac{E(z_{i+1}|g)}{g} \simeq \dfrac{g}{n}\left( 1-\dfrac{g}{n}\right) ^{i},
$ that becomes the exponential $\dfrac{1}{\chi }e^{-\frac{i}{\chi }},$ $\chi
=\frac{n}{g}$. This remark explains why large scale simulations of the BDY game with $n >> g$ 
appear compatible with an
exponential wealth distribution.

\subsection{Comparison with Monte Carlo simulations}
In this section, the results of Monte Carlo simulations are compared with the exact
equilibrium wealth distribution. The simulations have been performed
on a standard desktop computer equipped with a 1GHz processor. In the
initial state all the agents are given the same amount of coins. After an equilibration
run of 1000 MC steps, the values of $z_i$ have been sampled and
averaged over $10^5$ MC steps. In the cases reported in Figs. 1-3, the execution
time is a few seconds.

It is interesting to remark that for small values of $g$ the distribution is
strongly dependent on $g$: it is uniform for $g=2,$ linear for $g=3$,
parabolic for for $g=4$,$\ldots$. Except for the very peculiar case $g=2$, the
distribution is decreasing for $i>1$, but in some cases $E(z_{0}) < E(z_{1})$.
The latter feature deserves further investigations. 
Fig.1 shows the case $g=3$ and $n=3$, whereas Fig.2 has again 3 agents, but 
30 coins. Fig.3 is the
logarithmic graph for $g=30$ and $n=30$ to illustrate the approach to an
exponential-type distribution for large values of the number of agents, $g$.

%
\begin{figure}
\resizebox{0.75\columnwidth}{!}{%
\includegraphics{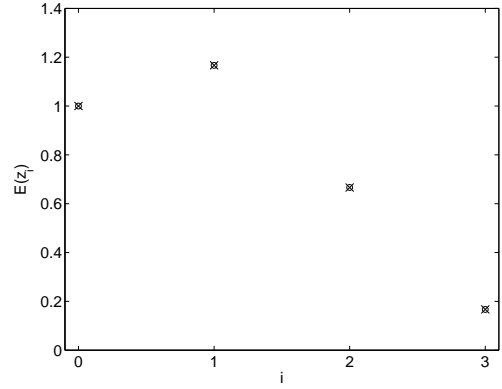}
}
\caption{Theoretical (cross) and simulated (circle) points for $g=3$, $n=3$, after $10^{5}$ simulation steps.}
\label{fig:1}       
\end{figure}

\begin{figure}
\resizebox{0.75\columnwidth}{!}{%
\includegraphics{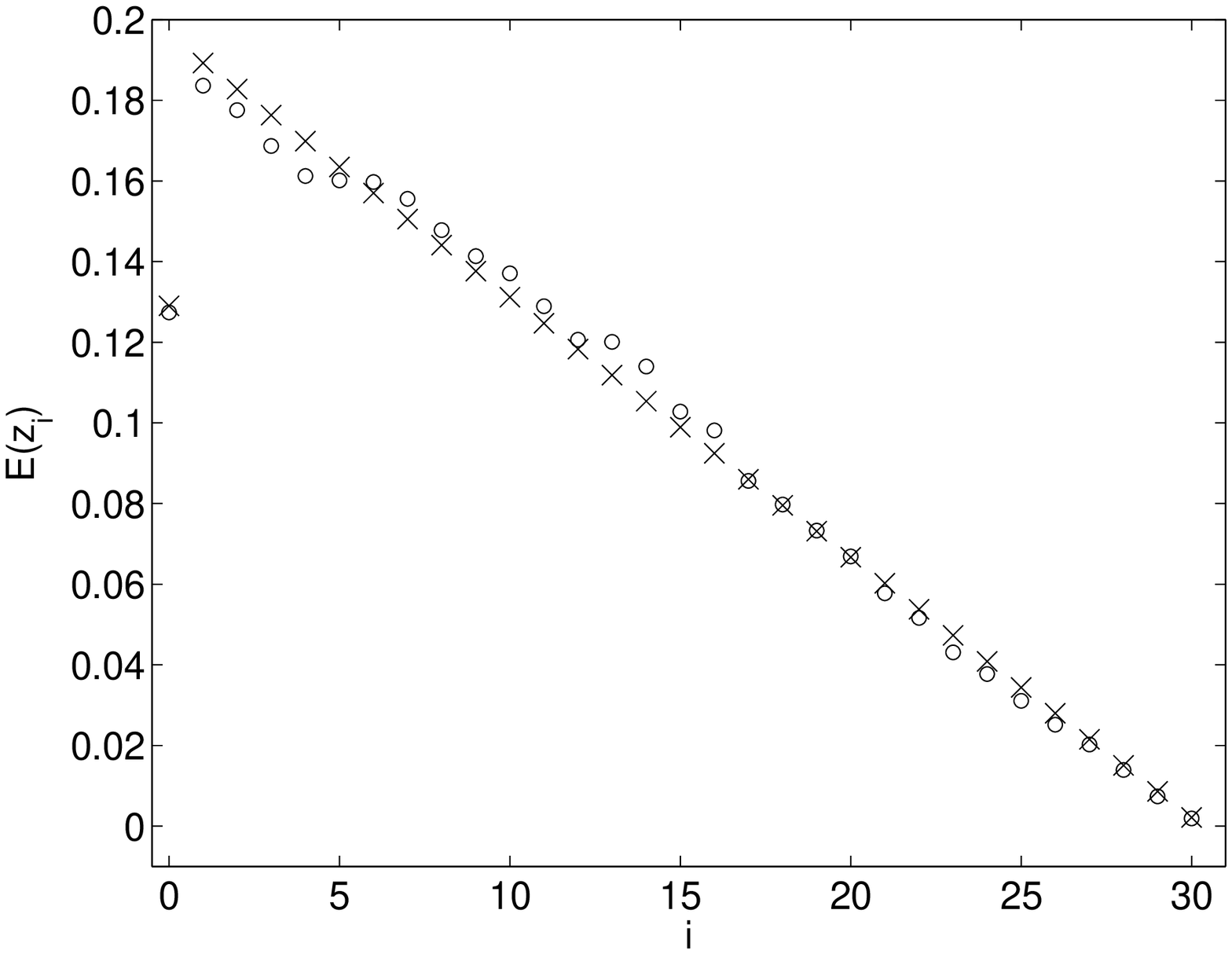}
}
\caption{Theoretical (cross) and simulated (circle) points for $g=3$, $n=30$, after $10^{5}$ simulation steps.}
\label{fig:2}       
\end{figure}

\begin{figure}
\resizebox{0.75\columnwidth}{!}{%
\includegraphics{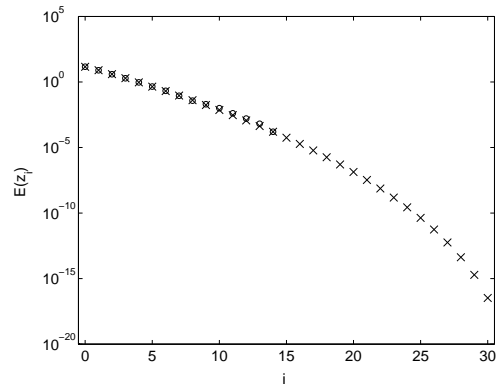}
}
\caption{Theoretical (cross) and
simulated (circle) points for $g=30$, $n=30$, after $10^{5}$ simulation steps. The simulation is
too short to reproduce the smaller values of $E(z_{i})$ for $i \geq 15$.}
\label{fig:3}       
\end{figure}

\section{Discussion and conclusions}
Recently, parsimonious exchange games like the one studied in this paper have been challenged by a group
of leading non-orthodox economists \cite{lux05,gallegati06}. These games have been introduced 
in order to explain the allocation of wealth in the presence of finite resources.  
In \cite{lux05,gallegati06}, they are considered unrealistic because they do not take into account the free will of agents to participate in an exchange, and
they include only strictly conserved resources, without production.
Incidentally, in games such as the OPIP or the BDY models, inequality is obtained by pure chance. Rich agents 
have no specific individual merit. Based on their beliefs, some scholars could also dislike this feature.

Replies to the objections in \cite{lux05,gallegati06} have already appeared in two papers by
Angle \cite{angle06a} and by McCauley \cite{mac06}. In particular, Angle presents various arguments in favour
of parsimonious exchange games, including their ability to reproduce empirical facts \cite{angle06b}.

The present authors would also like to stress that, also thanks to simple exchange models, 
a new concept of equilibrium could find its way into Economics: namely {\em Statistical equilibrium}.
Many stochastic models in Economics are Markov chains or Markov processes (see refs. 
\cite{silver02,foellmer05,bottazzi04} for recent examples) and the concepts developed in this paper
apply to those cases. These ideas will be the subject of future papers on the role
of statistical equilibrium in Economics. The reader can consult
ref. \cite{foley94} for an early discussion and refs. \cite{mac06,mac04} for a 
criticism on the relevance of thermodynamic equilibrium in Economics.

One of the main results of this paper is eq. (\ref{wealthdistribution}), giving the 
so-called {\em wealth distribution}. As the agent descriptions are not equiprobable,
previous statistical mechanical arguments have to be
revised. In general, the wealth distribution is not exponential and it becomes exponential
only in the appropriate limit of large density and large number of agents. It is interesting to study
the rate of approach to equilibrium in the BDY model, but this will the subject
of a future paper of this series. The next paper of the series, will be devoted to a set of 
simple exchange models for the redistribution of wealth that can be regarded as toy taxation
mechanisms.

\section*{APPENDIX}

\noindent \textbf{The normalization constant} \\

The total number of possible agent descriptions, $\mathbf{n}$, is 
$$
W(g,n)=%
\dbinom{n+g-1}{n},
$$
and they can be classified in terms of the number of
agents with at least one coin: $k=g-z_{0}$, $k=1,...,g$. Therefore, the number of
agent descriptions with $k$ {\em fixed} agents with at least one coin is given by all
occupation numbers which allocate $n-k$ coins to $k$ agents, that is 
$$%
\dbinom{n-k+k-1}{n-k}=\dbinom{n-1}{n-k}=\dbinom{n-1}{k-1},
$$ 
while $\dbinom{g}{k}$ are the
different ways to choose the $k$ agents among the $g$ available. Then 
$$%
W(k,g,n)=\dbinom{g}{k}\dbinom{n-1}{n-k}
$$ 
is the number of agent descriptions with $k$ agents with at least one coin. Indeed, one has:
$$%
\dbinom{n+g-1}{n}=\sum_{k=1}^{g} \dbinom{g}{k}\dbinom{n-1}{n-k},
$$
and this formula expresses the decomposition of all possible states in terms of their
\textquotedblleft support\textquotedblright\ $k$. The decomposition can be re-written as: 
$$%
W(g,n)=\sum_{k=1}^{g}W(k,g,n).
$$ 
Turning to eq.(\ref{inv}), the sum on all
states can be divided into a sum over $k$ and a sum over $\mathbf{n}|k$,
that is:
\begin{eqnarray}
1=\sum_{\mathbf{n}}\pi (\mathbf{n})=C\sum_{\mathbf{n}}k(\mathbf{n}%
)=C\sum_{k=1}^{g}k W(k,g,n)= & & \nonumber \\
C\sum_{k=1}^{g}k \dbinom{g}{k}\dbinom{n-1%
}{n-k}, \; \; \; \; \; \; \; \; \; \; \; \;  \; \; \; \;  & & \nonumber
\end{eqnarray}
which gives the desired normalization constant.\\

\noindent \textbf{Derivation of equation (}\ref{Ezik}\textbf{)} \\

The average number of agents whose occupation number is equal to $i$ is 
$$%
E(z_{i})=\sum_{j=1}^{g}P(Y_{j}=i)=gP(Y_{j}=i), 
$$ 
the last equality
holding as the $Y^{\prime }s$ are equidistributed. $P(Y_{j}=i)$, $i=0,1,...,n$
is the marginal equilibrium probability of the wealth of the $jth-$agent,
and it is the same for all $j$'s. 
It is necessary to study the marginal distribution of
an agent associated to the agent description probability (\ref{inv}) and to the
partition probability (\ref{invz}), both holding at equilibrium. In order to
derive formula \textbf{\ (}\ref{Ezik}\textbf{)}, one needs 
$$%
E(z_{i}|k)=gP(Y_{j}=i|k):
$$
the marginal wealth distribution of an agent
conditioned to $k = g - z_0$. One knows from (\ref{inv}) that all
agent descriptions $\mathbf{Y=n:=(}Y_{1}=n_{1},...,Y_{g}=n_{g})$ with the
same $k$ are equiprobable, and their number is 
$$W(k,g,n)=\dbinom{g}{k}%
\dbinom{n-1}{n-k}=\dbinom{g}{k}\dbinom{n-1}{k-1}.$$
Then $%
P(Y_{1}=n_{1}|k):=P(Y=i|k)$ is equal to the number of $\mathbf{Y}^{\prime }s$
such that $g-1$ agents share $n-i$ coins divided by $W(k,g,n).$ The
calculation can be divided into three parts. First, let us consider $P(Y=0|k)$; one has:
\begin{eqnarray}
P(Y=0|k)=\dfrac{W(k,g-1,n)}{W(k,g,n)} & = &
\frac{\dbinom{g-1}{k}\dbinom{n-1}{k-1}}{\dbinom{g}{k}\dbinom{n-1}{k-1}}= \nonumber \\
\frac{\dbinom{g-1}{k}}{\dbinom{g}{k}} = \frac{(g-1)!}{(g-1-k)!}\frac{(g-k)!}{g!} 
& = & \frac{g-k}{g}, \nonumber
\end{eqnarray}
then, let us consider $P(Y=i|k)$ with $k \geq 2$, and $i>0$; 
as there are $k-1$ agents left with at least one coin, one has:
\begin{eqnarray}
P(Y=i|k) & = & \dfrac{W(k-1,g-1,n-i)}{W(k,g,n)} = \; \; \; \; \; \; \; \; \nonumber \\
\frac{\dbinom{g-1}{k-1}\dbinom{%
n-i-1}{k-2}}{\dbinom{g}{k}\dbinom{n-1}{n-k}} & = & \dfrac{k}{g}\dfrac{\dbinom{n-i-1%
}{k-2}}{\dbinom{n-1}{k-1}},
\end{eqnarray}
finally, for $k=1$, one gets:
$P(Y=i|k=1)=\frac{\delta _{i,n}}{g},$ for $i>0$, as in this case all coins are
concentrated on a single agent. Eventually, by determining $E(z_{i}|k)$, one obtains eq. (\ref{Ezik}%
) as required.

\section*{ACKNOWLEDGEMENTS}

E.S. acknowledges useful discussion with Giulio Bottazzi, Mauro Gallegati, Eric Guerci, David Mas, Marco Raberto, and
Alessandra Tedeschi during a Thematic Institute sponsored by the Exystence EU network held in Ancona in 2005. 
He is grateful to J. Angle and J. McCauley
for pointing him to refs. \cite{angle06a} and \cite{mac06}, respectively. This work 
has been partially supported by MIUR project "Dinamica di altissima
frequenza nei mercati finanziari".


\begin{thebibliography}{9}

\bibitem{liggett85}
T.M. Liggett, \textit{Interacting Particle Systems}, (Springer, Berlin, 1985).

\bibitem{angle86}
J. Angle, 
\textit{The Surplus Theory of Social Stratification and the Size Distribution of 
Personal Wealth}, 
Social Forces {\bf 65}, 293--326, (1986).

\bibitem{angle96}
J. Angle, 
\textit{How the Gamma Law of Income Distribution Appears Invariant under Aggregation}, 
Journal of Mathematical Sociology {\bf 21}, 325--358, (1996).

\bibitem{angle02}
J. Angle, 
\textit{The statistical signature of pervasive competition on wages and salaries}, 
Journal of Mathematical Sociology {\bf 26}, 217--270, (2002).

\bibitem{bennati88}
E. Bennati,
\textit{Un metodo di simulazione statistica per l'analisi della distribuzione del reddito}, 
Rivista Internazionale di Scienze Economiche e Commerciali {\bf 35}, 735--756, (1988).

\bibitem{bennati93}
E. Bennati,
\textit{Il metodo di Montecarlo nell'analisi economica}, 
Rassegna di Lavori dell'ISCO, Anno X, n. 4, 31--79, (1993).

\bibitem{dragulescu00}
A. Dragulescu and V. M. Yakovenko, \textit{Statistical mechanics of money}, 
Eur. Phys. J. B {\bf 17}, 723--729 (2000).

\bibitem{costantini}
D. Costantini and U. Garibaldi,
\textit{The Ehrenfest Fleas: from Model to Theory}, Synthese
{\bf 139}, 107-142, (2004).

\bibitem{boltzmann68}
L. Boltzmann,
\textit{Studien \"uber das Gleichgewicht der lebendingen Kraft zwischen bewegten materielle Punkten} (1868) 
in \textit{Wissenschaftliche Abhandlungen}, vol. I, F. Hasenh\"orl (ed.), Leipzig, Barth, pp. 49-96 (1909).

\bibitem{bach90}
A. Bach, \textit{Boltzmann's Probability Distribution of 1877}, Archive for History of Exact Sciences 
{\bf 41}, 1--40, (1990).

\bibitem{costantini97}
D. Costantini and U. Garibaldi,
\textit{A Probabilistic Foundation of Elementary Particle Statistics. Part I}
Stud. Hist. Phil. Mod. Phys. {\bf 28}, 483--506, (1997).

\bibitem{hill77}
B.M.Hill, \textit{The Rank-Frequency Form of Zipf's Law, }%
Jour. Am. Stat. Ass. \textbf{69 }(348), 1017-1026, (1977).


\bibitem{lux05}
T. Lux, 
\textit{Emergent statistical wealth distributions in simple monetary exchange models: A critical 
review}, in \textit{Econophysics of Wealth Distribution}, A. Chatterjee, S. Yarlagadda, 
B.K. Chakrabarti eds., (Springer, Berlin, 2005).

\bibitem{gallegati06}
M. Gallegati, S. Keen, T. Lux, P. Ormerod,
\textit{Worrying Trends in Econophysics},
working paper, 2006.

\bibitem{angle06a}
J. Angle,
\textit{A Comment on Gallegati et al.'s ``Worrying Trends in Econophysics''},
working paper, 2006.

\bibitem{mac06}
J. McCauley,
\textit{Response to ``Worrying Trends in Econophysics''}
working paper, 2006.

\bibitem{angle06b}
J. Angle,
\textit{The Inequality Process as a wealth maximizing process}
Physica A, {\bf 367}, 388--414 (2006).

\bibitem{silver02}
J. Silver, E. Slud, and K. Takamoto,
\textit{Statistical Equilibrium Wealth Distributions in an
Exchange Economy with Stochastic Preferences}
Journal of Economic Theory {\bf 106}, 417--435 (2002).

\bibitem{foellmer05}
H. F\"ollmer, U. Horst, and A. Kirman,
\textit{Equilibria in financial markets with heterogeneous agents: A probabilistic
perspective},
Journal of Mathematical Economics {\bf 41}, 123--155 (2005).

\bibitem{bottazzi04}
G. Bottazzi, G. Dosi, G. Fagiolo, and A. Secchi,
\textit{Sectoral and Geographical Specificities in the Spatial Structure of Economic Activities}
Scuola Superiore S.Anna, LEM Working Paper 2004/21, (2004).

\bibitem{foley94}
D.K. Foley,
\textit{A statistical equilibrium theory of markets}, Journal of 
Economic Theory {\bf 62}, 321--345 (1994).

\bibitem{mac04}
J. McCauley,
\textit{Dynamics of Markets: Econophysics and Finance}, (Cambridge University Press, Cambridge UK, 2004).

\end{thebibliography}
\end{document}